\documentclass[12pt,preprint]{aastex}
\def\lsim{\lower.5ex\hbox{$\; \buildrel < \over \sim \;$}}
\def\gsim{\lower.5ex\hbox{$\; \buildrel > \over \sim \;$}}
\lefthead{}
\righthead{  }

\def\lsim{\lower.5ex\hbox{$\; \buildrel < \over \sim \;$}}
\def\gsim{\lower.5ex\hbox{$\; \buildrel > \over \sim \;$}}

\received{2002 September 8}
\begin{document}

\title{Growing hydrodynamic modes in Keplerian accretion disks during secondary perturbations:
Elliptical vortex effects}

\author{Banibrata Mukhopadhyay\altaffilmark{!}}
\altaffiltext{!}{bmukhopa@cfa.harvard.edu, bm@physics.iisc.ernet.in}

\affil{1. Institute for Theory and Computation, Harvard-Smithsonian Center for Astrophysics,
60 Garden Street, MS-51, Cambridge, MA 02138, USA\\
2. Department of Physics, Indian Institute of Science, Bangalore-560012, India}


\begin{abstract}

The origin of hydrodynamic turbulence, and in particular of an anomalously
enhanced angular momentum transport, in accretion disks is still an unsolved
problem.
This is especially important for cold disk systems which are practically neutral in
charge and therefore turbulence can not be of magnetohydrodynamic origin.
While the flow must exhibit some instability and then 
turbulence in support of the transfer of mass inward
and angular momentum outward, according to the linear perturbation theory, in absence
of magnetohydrodynamic effects, it
should always be stable. We demonstrate that the three-dimensional
secondary disturbance to the primarily perturbed disk, consisting of elliptical vortices, gives
significantly large hydrodynamic growth in such a system and hence may 
suggest a transition to an ultimately turbulent state. 
This result is essentially applicable to accretion disks
around quiescent cataclysmic variables, in proto-planetary and star-forming disks,
the outer region of disks in active galactic nuclei, where the gas is significantly cold
and thus the magnetic Reynolds number is smaller than $10^4$.

\end{abstract}

\keywords{accretion, accretion disks --- hydrodynamics --- turbulence --- instabilities}

\section{Introduction}

Origin of turbulence and then the viscosity parameter $\alpha$, 
while has been demonstrated for a hot Keplerian accretion disk by the
magnetorotational instability (MRI; Balbus \& Hawley 1991; Balbus, Hawley, \&
Stone 1996; Hawley, Balbus, \& Winters 1999), is
still not well understood for cold disks, e.g. accretion disks
around quiescent cataclysmic variables (Gammie \& Menou 1998; Menou 2000),
proto-planetary and star-forming disks (Blaes \& Balbus 1994), and
the outer region of disks in active galactic nuclei (Menou \&
Quataert 2001), even after more than three decades of the famous discovery of
the $\alpha$-disk (Shakura \& Sunyaev 1973). 
These cold systems are largely neutral in charge
so that the MRI driven turbulence appears to be ruled out.
However, to support accretion, there must appear some sort of turbulence (most likely
pure hydrodynamic in nature) and then the corresponding turbulent viscosity, as
the molecular viscosity is negligible. Richard \& Zahn (1999) indeed showed 
by analyzing results of laboratory experiments for the classical Couette-Taylor
flow that in the case where angular momentum increases outward, similar to the
Keplerian flow, hydrodynamic turbulence may be sustained. Then they derived the 
corresponding turbulent viscosity which is very useful for the astrophysical purpose.
Longaretti (2002) argued that the Keplerian
accretion flow in the shearing sheet limit should be turbulent and
the lack of turbulence in simulations is due to lack of their resolution.
Then Lesur \& Longaretti (2005) showed that the required resolution is not
possible to achieve with the presently available computer resources.
They also found that the efficiency of turbulent transport is directly correlated
to the critical Reynolds number for the transition to turbulence.
Inspite of all these analyzes, the actual origin of hydrodynamic turbulence in such systems
still remains unclear. It is important to note that the numerical simulation has never been carried out 
with a {\it very high} Reynolds number what the real accretion disk exhibits. Therefore,
the hydrodynamic effects may compete with the magnetohydrodynamic effects (essentially MRI) 
at a high Reynolds number and, in a realistic system, turbulence may be 
due to hydrodynamic effects independent
of whether the disk is cold or hot.

Recently, several authors including ourselves have
put their efforts in achieving progress toward the solution of this
difficult problem (e.g. Tevzadze et al. 2003; Chagelishvili 2003;
Umurhan \& Regev 2004; Yecko 2004; Mukhopadhyay, Afshordi \& Narayan 2005; Afshordi, Mukhopadhyay
\& Narayan 2005; Johnson \& Gammie 2005; Umurhan et al. 2006; 
Barranco \& Marcus 2005; Papaloizou 2005). 
The main aim of these works is to demonstrate the
{\it pure} hydrodynamic turbulence by transient growth of energy 
with the suitable choice of two-dimensional initial perturbation. The idea is any large growth
plausibly could switch the accretion disk into the nonlinear regime which might
result its subcritical transition to turbulence
if energy growth exceeds the threshold for turbulence.
One might argue that transient growth, even if large, can not provide much
clue on the existence and properties of a turbulent basin of attraction.
Schmid \& Henningson (2001) and Criminale, Jackson \& Joslin (2003) have 
described in detail that transition to turbulence is not a unique process
but it depends on the initial condition/disturbance and the nature 
of the flow. In fact, it is known that even in presence of secondary instability
linearly unstable base flows may reach to a non-turbulent saturated state.
However, turbulence definitely belongs to the nonlinear regime and it 
is exhibited only in the situations when large growth (or more precisely transient
growth for the present purpose) switches the system over the non-linear regime.
As our present goal is to understand the possible origin of hydrodynamic turbulence,
we consider those situations when large transient growth governs non-linearity.
At the present case where the accretion time scale is comparable
(or shorter) to the time scale to achieve maximum growth (MAN), decaying
nature of growth
at a period longer than the accretion time scale does not matter. If growth (or maximum growth) 
becomes large enough that exceeds the
threshold to trigger non-linearity and 
turbulence, then it does not matter whether growth evolves exponentially or transiently.

Dauchot \& Manneville (1997) argued with a toy model that transient growth
does not guaranty the transition to turbulence as the underlying phase portraits
do not get deformed much in absence of any linearly unstable mode. However, that does not
rule out the importance of transient growth as their model is very 
simplistic consisting of two variables only and growth is tinny to 
expect any deformation in the phase portraits. Other alternative methods were
proposed to describe the subcritical transition to turbulence and then
to investigate various underlying aspects in detail by e.g. Waleffe (1995, 1997),
Brosa \& Grossmann (1999), Waleffe (2003), Kerswell (2005), while last two works,
that discuss non-rotating Couette flows and pipe flows respectively,
are mostly relevant for the astrophysical purpose. Although the works bring some new 
insight into the subject, the main problem i.e. to understand the subcritical turbulence
in three-dimension remains unsolved. This is indeed a non-trivial problem not
only in astrophysics but also in fluid dynamics. In the present paper, we suggest
a possible mechanism to govern non-linear effect into a Keplerian accretion disk.
While our prescription does not solve the problem completely, it certainly
opens up a new avenue to understand the physics behind this puzzle.

The problem with transient growth, what has been proposed as the mechanism to 
generate turbulence
(e.g. Umurhan \& Regev 2004; hereafter UR; Mukhopadhyay, Afshordi \& Narayan 2005,
Afshordi, Mukhopadhyay \& Narayan 2005; hereafter MAN), is that in two-dimension
underlying perturbations  
must ultimately decline to zero 
in the viscous flow. To overcome this limitation, it is necessary to invoke
three-dimensional effects.  Various kinds of secondary
instability, such as the elliptical instability, are widely
discussed as a possible route to self-sustained turbulence in
linearly perturbed shear flows (see, e.g. Pierrehumbert 1986;
Bayly 1986; Hellberg \&
Orszag 1988; Craik 1989; Le Dize\'s, Rossi \& Moffatt 1996;
Kerswell 2002). Therefore, we are motivated to see
whether these three-dimensional instabilities are present in
the Keplerian flows
which consist of elliptical streamlines under two-dimensional perturbation.
Goodman (1993) first explored the
possible role of the elliptical instability in an accretion disk
and Lubow, Pringle \& Kerswell (1993) and Ryu \& Goodman (1994)
showed that angular momentum may be transferred from the disk to
the tidal source by the instability effect. They essentially
considered the case of forced disks. Later, Ioannou \&
Kakouris (2001) also examined transient growth being constantly re-excited
by external noise.

The three-dimensional instability of a two-dimensional
flow with elliptical vortices has been demonstrated by number of authors
(e.g. Craik \& Criminale 1986; Waleffe 1990) and
has been proposed as a generic mechanism for the breakdown of many two-dimensional
high Reynolds number flows. 
This result motivates us to investigate whether the similar
scenarios exist in a Keplerian accretion disk.
Therefore, essentially we plan to investigate the
Keplerian flow with two consecutive perturbation effects. The primary two-dimensional perturbation,
which drives transient growth (MAN), governs elliptical streamlines in the flow.
Then we consider the further three-dimensional perturbation,
that is the secondary one, to this two-dimensional flow, if that drives 
three-dimensional instability effect
in presence of viscosity. Presumably, three-dimensional instabilities
lead to nonlinear feedback and self-sustained turbulence.

Although the maximum growth in the Keplerian disk with finite vertical thickness and then with
finite vertical perturbation is smaller compared to that in a two-dimensional Keplerian
disk, as Tevzadze et al. 
(2003) and MAN argued, vertical stratification may cause to have
a nonvanishing asymptotic growth that is a fraction of the maximum growth. However, they could
not show whether this will significantly help the onset of turbulence.
By three-dimensional hydrodynamic simulations, Barranco \& Marcus (2005) studied
the dynamics and formation of vortices in stably stratified proto-planetary disks
and found that vortices are unstable under perturbation. Theoretically, the origin of these
vortices is understood as perturbation in plane shear flow 
(UR, MAN). Continuing in the same line of thought, recently, 
Umurhan et al. (2006) have shown substantial growth resulting
from initial perturbations in the linearly stable steady state and have concluded
that significant perturbation energy amplification occurs in accretion disks on a global scale.

In the present paper, we perform a local linear 
analysis in the shearing box approximation.
Although a small section of an accretion disk may not reproduce the global
disk properties, we assume that if turbulence is exhibited at any section,
then that sustains and eventually affects the entire disk. While the local
result does not guaranty its global signature, absence of local instability
and then possible turbulence, perhaps, does guaranty its global absence. 
Therefore, from the results of this paper we can determine the future 
avenues of the subject, whether the analyzes for
secondary instability including transient growth 
is a fruitful path to solve the problem.

In the next section, we outline the basic model considered for this 
problem describing primary and secondary perturbation. 
In \S3, we present the simple solution when the evolution of
secondary perturbation is much rapid than that of primary one. Subsequently, in \S4 we analyze the
general solutions when both perturbations vary simultaneously. Finally we sum up the results
with a discussion in \S5.

\section{The Model}



We consider a small portion of the Keplerian flow centered on radius $r_0$ (see MAN for detail).
Therefore, the Keplerian flow locally reduces to rotating Couette flow in a narrow gap limit
whose unperturbed velocity vector is $\vec{U}=(0,-x,0)$, where $x<<r_0$. 
Therefore, the Navier-Stokes and the continuity equations for
the dynamics of viscous incompressible disk fluid are
given by 
\begin{equation}
\frac{\partial \vec{U}}{\partial t}+\vec{U}.\nabla\vec{U}+\vec{\Omega}
\times\vec{\Omega}\times \hat{x}\,x
+2\vec{\Omega}\times\vec{U}+\nabla(\tilde{p})=\frac{1}{R}\nabla^2\vec{U};\,\,\,\, \nabla.\vec{U}=0,
\label{nevst2}
\end{equation}
where $\nabla=(\partial/{\partial x},\partial/{\partial y},\partial/{\partial z})$,
$R$ is the Reynolds number, 
the angular frequency $\vec{\Omega}=(0,0,1/q)$,
$t$ is the time, and $\tilde{p}$ is proportional
to the pressure. For a Keplerian disk $q=1.5$ and for a constant angular
momentum disk $q=2$. 
Here, all the variables are expressed in dimensionless units.
The unit of length is the box size in the radial direction and the unit of velocity
is the maximum relative velocity between two fluid elements into the box.
Below we describe two subsequent perturbation effects, primary and secondary, one by one.

\subsection{Primary Perturbation}

Under a linear two-dimensional (primary) 
perturbation, the velocity vector of the (primary) flow reduces as
\begin{equation}
\vec{U}\rightarrow\vec{U}_p=(w_x,-x+w_y,0)={\bf A}.\vec{d},
\label{primper}
\end{equation}
where
\begin{equation}
w_x=\zeta\frac{k_y}{\kappa^2}\sin(k_xx+k_yy),\,\,w_y=-\zeta\frac{k_x}{\kappa^2}\sin(k_xx+k_yy),
\label{primpert}
\end{equation}
${\bf A}$ is a tensor of rank $2$, the position vector $\vec{d}=(x,y,z)$,
$k_x,k_y$ are the components of the wave vector of perturbation,
$\kappa=\sqrt{k_x^2+k_y^2}$, and $\zeta$ is the amplitude of the vorticity perturbation.
Now we concentrate on a further small patch of the primarily perturbed flow so that the spatial scale
is very small compared to the wavelength of primary perturbation, $x<<1/k_x$,
$y<<1/k_y$. Therefore, $\bf A$ comes out to be
\begin{equation}
{\bf A}=A_j^k=\left(\begin{array}{ccc} \zeta\sqrt{\epsilon(1-\epsilon)} &
\zeta(1-\epsilon) & 0\\ -(1+\zeta\epsilon) &
-\zeta\sqrt{\epsilon(1-\epsilon)} & 0\\ 0 & 0 & 0 \end{array}\right),
\label{axy}
\end{equation}
where $\epsilon=(k_x/\kappa)^2$.
The above $\bf A$ indicates the flow pattern to be a generalized elliptical one compared
to that discussed in standard fluid literature (e.g. Bayly 1986; Craik 1989; Kerswell 2000)
as an ordinary elliptical flow given by
\begin{equation}
A_j^k=\left(\begin{array}{crr} 0 & 1-\epsilon & 0\\ -(1+\epsilon) & 0 &
0\\ 0 & 0 & 0 \end{array}\right).
\label{axy0}
\end{equation}

\subsection{Secondary Perturbation}

The background flow for further perturbation (secondary one) 
also corresponds to eqn. (\ref{nevst2}) except
$\vec{U}$ is replaced by $\vec{U}_p$.
The secondary perturbation modifies components of the velocity
in eqn. (\ref{primper}) and the pressure as $\vec{U}_p\rightarrow \vec{U}_p+\vec{u}$ and
$\tilde{p}\rightarrow\tilde{p}+p$. The perturbation is considered to be
plane wave typed given by
\begin{equation}
(u_i,\tilde{p})=(v_i(t),p(t)) \exp(ik_m(t) x^m),
\label{pet}
\end{equation}
where $k_m>>k_x,k_y$. The Latin indices run from $1$ to $3$ such that
e.g. $x^m\equiv (x,y,z)$, and thus the background velocity
can be written as $U_{pi}=A^m_i\,x_m$.
Therefore, from eqns. (\ref{nevst2}), (\ref{axy}) and (\ref{pet}) with replacing $\vec{U}$
by $\vec{U}_p$
and after some algebra, we obtain the evolution of a linear secondary perturbation
\begin{equation}
\dot{v}_j+A_j^k\,v_k+2\,\epsilon_{mkj}\Omega^m v^k=-ip\,k_j
-\frac{v_j}{R}\,k^2,
\label{perteq}
\end{equation}
along with
\begin{equation}
k_nv^n=0,\,\,\,\,\,
\dot{k}_j=-(A^m_j)^T\,k_m,\,\,\,\,\,
k_n\dot{v}^n=k_m\,A^m_n\,v^n,
\label{keq}
\end{equation}
where the `over-dot' indicates a derivative with respect to $t$,
$\epsilon_{mkj}$ is a Levi-Civita tensor, and $k^2=k_mk^m$.

Two components, $k_1$ and $k_2$, of the wave-vector [$\vec{k}=k_m=(k_1,k_2,k_3)$] of
secondary perturbation oscillate in time with the angular frequency
$\varpi=\sqrt{\zeta(1-\epsilon)}$ at a fixed $\epsilon$, while the third one, $k_3$, remains 
constant. As we choose the signature of the background Minkowski space-time to be $[-,+,+,+]$,
it does not matter whether
any Latin index appears as a lower case or an upper case. For example,
$A^k_j=A_{jk}$, where $j$ and $k$ respectively indicate the row and the column
number.

Now projecting out eqn. (\ref{perteq}) by $P^j_i=\delta^j_i-k^{-2}k^j\,k_i$ and
using eqn. (\ref{keq}) we obtain
\begin{equation}
\dot{v}_i=\left(2\frac{k^j\,k_i}{k^2}-\delta^j_i\right)A^k_j\,v_k
-2\,\epsilon_{mki}\,\Omega^m\,v^k-\frac{v_i}{R}k^2
+\left(2\,\epsilon_{mkj}\,\Omega^m\,v^k+\frac{v_j}{R}k^2\right)\frac{k^jk_i}{k^2}.
\label{veq}
\end{equation}
A similar equation was obtained by Bayly (1986), except that they now have additional
terms induced due to the Coriolis and the viscous effects.

As $R$ is very large in an accretion disk,
we neglect the viscous term in eqn. (\ref{veq}) comparing with others and
rewrite the equation as
\begin{equation}
\dot{v}_i=\Lambda^j_i\,v_j,
\label{vmat}
\end{equation}
where
\begin{equation}
\Lambda_{ij}=\left(\begin{array}{ccc} \left(\frac{2k_1^2}{k^2}-1\right)A_{11}+
\frac{2k_1k_2}{k^2}\left(A_{21}+\frac{1}
{q}\right) & \frac{2k_1^2}{k^2}\left(A_{12}-\frac{1}{q}\right)+\frac{2}{q}-A_{12}+\frac{2k_1k_2}{k^2}A_{22}
 & 0\\ \frac{2k_1k_2}{k^2}A_{11}+\frac{2k_2^2}{k^2}\left(A_{21}+\frac{1}{q}\right)-A_{21}-\frac{2}{q}
 & \frac{2k_1k_2}{k^2}\left(A_{12}-\frac{1}{q}\right)+\left(\frac{2k_2^2}{k^2}-1\right)A_{22} &
0\\ \frac{2k_1k_3}{k^2}A_{11}+\frac{2k_2k_3}{k^2}\left(A_{21}+\frac{1}{q}\right) &
\frac{2k_1k_3}{k^2}\left(A_{12}-\frac{1}{q}\right)+\frac{2k_2k_3}{k^2}A_{22}  & 0 \end{array}\right).
\label{mat}
\end{equation}
We essentially need to solve eqn. (\ref{vmat}) to demonstrate the behavior of perturbations.
Next we describe the solution in various possible situations.

\section{Perturbation Solution at a fixed $\epsilon$}

The solution with constant $\epsilon$ 
\footnote{This is to remind that
the radial wave-vector of primary perturbation, $k_x$, varies with $t$ (MAN),
therefore in reality the background for secondary perturbation appears to be time-dependent.}
implies the result at a particular instant of the evolution of primary perturbation
that provides the {\it instantaneous growth rate}. 
The underlying idea is to focus on
the situation when the evolution of secondary perturbation and the corresponding 
development of growth is much rapid compared to that due to primary perturbation and therefore 
$\epsilon$ practically remains constant during secondary perturbation evolution.
This helps us to compare the result with what exists in standard fluid literature 
(e.g. Bayly 1986; Kerswell 2002). 

The general solution of
eqn. (\ref{vmat}) can be written as a linear superposition of Floquet modes
\begin{equation}
v_i(t)=\exp(\sigma\,t)\,f_i(\phi),
\label{flo}
\end{equation}
where $\phi=\varpi\,t$, $f_i(\phi)$ is a periodic function with
time-period $T=2\pi/\varpi$, and $\sigma$ is the Floquet exponent
which is the eigenvalue of the problem. Clearly, if $\sigma$ is positive then
perturbation increases with time.
Applying the periodicity condition, $f_i(2\pi)=f_i(0)$,
in eqn. (\ref{flo}), we obtain
\begin{equation}
v_i(T)=\exp(\sigma\,T)\,f_i(2\pi)=\exp(\sigma\,T)\,f_i(0)=\exp(\sigma\,T)\,v_i(0).
\label{veigen}
\end{equation}
Therefore, $\exp(\sigma\,T)$ serves as an evolution operator. 

To determine the exponential growth rate, $2\sigma$, we
strictly follow e.g. Bayly (1986) and Craik (1989). In this method,
one has to evaluate the associated velocity evolution matrix,
whose eigenvalue and eigenvector
at $t=T$ are $e^{\sigma T}$ and $f_i(2\pi)=f_i(0)$ respectively, satisfying
\begin{equation}
\frac{dM_{ji}(t)}{dt}=\Lambda^m_jM_{mi}(t),
\label{matevo}
\end{equation}
where $M_{ji}(0)=\delta_{ji}$. Essentially $M_{ji}(t)$ serves as an evolution operator
such that 
\begin{equation}
v_j(t)=M_{ji}(t)v_i(0). 
\label{vevo}
\end{equation}
Thus, using the fourth order Runge-Kutta
method, one can easily compute the elements of the $3\times 3$ matrix $M_{ji}(T)$.
Interesting feature to note is that $<Tr(\Lambda_{ij})>=0$ \footnote{`$Tr$'
denotes the sum of diagonal elements of the matrix, $<...>$ indicates
the averaged value, and `det' refers to the determinant of the matrix.}
over $0\le t\le T$. Therefore, $\det(M_{ji}(T))=1$.
Moreover, $d(k_jM^j_i(t))/dt=0$ and therefore $k_i(0)=k^j(0)M_{ji}(T)$ which
indicates that one eigenvalue of this $3\times 3$ matrix is always unity.
The remaining two eigenvalues of $M_{ji}(T)$ must be either real and
reciprocal to each other or complex conjugate to each other with unit modulus.
This property helps us to check the accuracy of results such that the
product of all three eigenvalues must be unity.
Therefore, our problem reduces to evaluate two no-trivial
eigenvalues, $\mu_1,\mu_2$, of the
matrix $M_{ji}(T)$. If $\mu_1$ or $\mu_2$ is real and positive, then the Floquet exponent
$\sigma_i=log(\mu_i)/T$.

When the initial value of $\vec{k}$, i.e., $\vec{k}(0)=k_m(0)\equiv(0,0,1)$,
$\vec{k}(t)$ remains conserved throughout as follows from eqn. (\ref{keq}). 
This is the {\it pure} vertical perturbation.
In this case, $\Lambda_{ji}$ in eqn. (\ref{mat}) appears
as a constant matrix.
Therefore, $e^{\sigma t}$ and $f_i(t)$ are
the eigenvalue and the eigenvector respectively of the matrix $M_{ji}(t)$ at any time $t$ and 
thus the Floquet exponent can be evaluated at any instant.
In fact, there is an exact analytical solution for
the Floquet exponents for this initial condition
which are the eigenvalues of the matrix $\Lambda_{ji}$. 

\subsection{Ideal elliptical flow}

Before investigating our accretion disk solutions in detail, let us recapitulate the nature of standard
elliptical flow and the corresponding instability discussed in the fluid dynamics
literature since long time. The velocity of a two-dimensional fluid element with elliptical
streamlines is given by 
$\vec{U}_p={\bf A}{\bf .}\vec{d}$ (see, e.g. Kerswell 2002) with the definition
of ${\bf A}$ as follows from eqn. (\ref{axy0}).

\begin{figure}
\epsscale{0.8}
\plotone{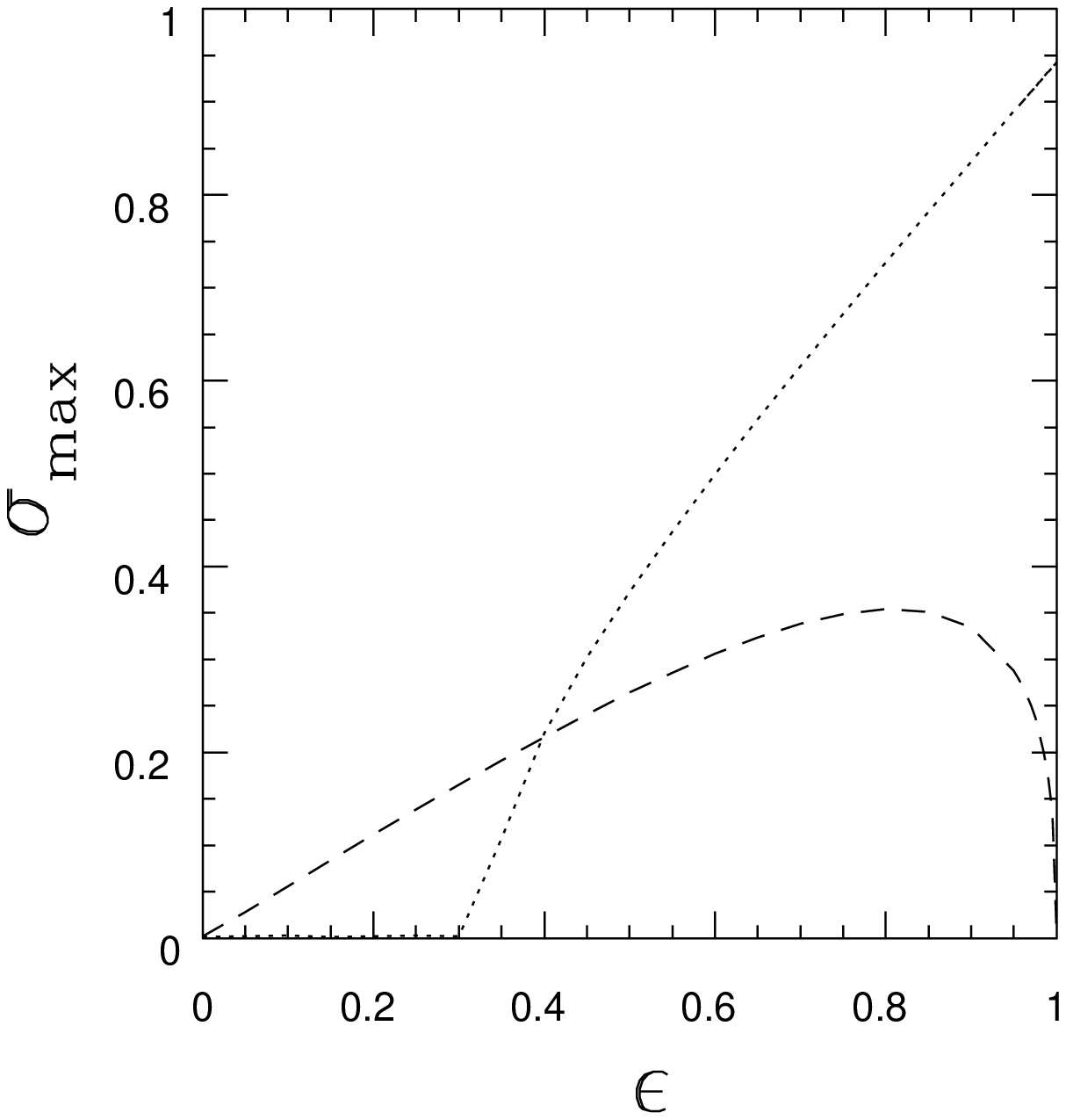}
\caption{ \label{ellip} Variation of maximized growth rate for
elliptical instability as a function of eccentricity parameter.
Dashed and dotted curves correspond to results for non-rotating
and rotating shear flows respectively. $q=1.5$. }
\end{figure}

Now following the perturbation technique described in \S2.2, we obtain
$\Lambda_{ij}$ given by eqn. (\ref{mat}) with the components of $\bf A$ as follows from
eqn. (\ref{axy0}).  If $\vec{k}$ is constant, i.e. perturbation is vertical, then $\Lambda_{ij}$
is also a constant matrix given by
\begin{equation}
\Lambda_{ji}=\left(\begin{array}{ccc} 0 & 2\Omega_z+\epsilon-1
 & 0\\ -2\Omega_z+\epsilon+1 & 0  &
0\\ 0 & 0 & 0 \end{array}\right),
\label{mat2}
\end{equation}
whose non-trivial eigenvalues are the velocity growth rates, Floquet exponents, given by
\begin{equation}
\sigma=\pm\sqrt{{\epsilon}^2-(1-2\Omega_z)^2}.
\label{eliec}
\end{equation}
For a Keplerian flow, $\sigma=\pm\sqrt{\epsilon^2-1/9}$. Therefore, the vertical
perturbation gives rise to the positive {\it instantaneous} growth rate in a Keplerian disk for
$\epsilon >1/3$. For any other perturbation, $k_1$ and $k_2$ oscillate in time
with the angular frequency $\varpi=\sqrt{1-\epsilon^2}$ and $k_3$ remains constant
(for detailed descriptions, see, e.g. Bayly 1986; Craik 1989; Kerswell 2002).
In Fig. \ref{ellip}, we compare the variation of maximum velocity growth rate as a function of
eccentricity parameter for a Keplerian flow with that of non-rotating flow. By ``maximum''
we refer the quantity obtained by maximizing over $\vec{k}$. Clearly the growth rate
is significantly large at high $\epsilon$ for a Keplerian system that is interesting for
the astrophysical purpose \footnote{This is to remind that our rotating Keplerian 
flow described in \S2 is
highly eccentric at large $k_x$, i.e. at the early stage of the evolution of primary perturbation.}.
This is understood physically from eqn. (\ref{eliec}).
When $\Omega_z=0$, from
eqn. (\ref{eliec}), $\sigma=0$ at $\epsilon=1$. However, for a rotating flow,
$\sigma=\pm 2\sqrt{q-1}/q$ at $\epsilon=1$. This result motivates us to study the elliptical
streamline (vortex) effects in an actual Keplerian flow governs in accretion disks 
as follows from eqn. (\ref{axy}).

\subsection{Flow in a Keplerian disk}

Here the velocity, $\vec{U}_p$, of the background flow (primarily perturbed flow) is defined according 
to $\bf A$ given by eqn. (\ref{axy}).
When does our primarily perturbed Keplerian flow 
reduce to conventional flows?
(1) When $k_x\rightarrow\infty$ i.e. $\epsilon\rightarrow 1$ and $\zeta\rightarrow 1$, 
$\bf A$ in eqn. (\ref{axy}) is same as that in eqn. (\ref{axy0}).
This is the case of an extremally eccentric flow.
(2) When $k_x\rightarrow\infty$ and $\zeta\rightarrow 0$, ${\bf A}$ in eqn. (\ref{axy}) reduces to
that of plane shear flow (MAN).
(3) When $k_x\rightarrow 0$ i.e. $\epsilon \rightarrow 0$ and $\zeta\rightarrow 1$, 
again the form of $\bf A$ in both the equations are same. This is the case of a circular flow.
(4) When $k_x\rightarrow 0$ and $\zeta\rightarrow 0$, ${\bf A}$ in eqn. (\ref{axy}) again reduces to
that of plane shear flow (MAN).

Now for a pure vertical perturbation, $\Lambda_{ji}$ in eqn. (\ref{mat}) reduces to
\begin{equation}
\Lambda_{ji}=\left(\begin{array}{ccc} -\zeta\sqrt{\epsilon(1-\epsilon)} & 2\Omega_z-\zeta(1-\epsilon)
 & 0\\ -2\Omega_z+\zeta\epsilon+1 & \zeta\sqrt{\epsilon(1-\epsilon)}  &
0\\ 0 & 0 & 0 \end{array}\right),
\label{mat1}
\end{equation}
where $\Omega_z=1/q$. As before the above $\Lambda_{ji}$ is a constant matrix, and thus we evaluate
the Floquet exponent as
\begin{equation}
\sigma=\pm\sqrt{\zeta\epsilon-(2\Omega_z-1)(2\Omega_z-\zeta)}.
\label{sigcon}
\end{equation}
Most of the works in fluid literature, so far,
have been carried out for non-rotating systems without focusing on
a Keplerian flow rigorously. Therefore, eqn. (\ref{sigcon}) can be seen as 
an extension of those results for an actual rotating Keplerian flow.

Now we understand the following points from eqn. (\ref{sigcon}).
(1) When $\Omega_3=0$, $\sigma=\sqrt{\zeta(\epsilon-1)}$. This verifies
that non-rotating two-dimensional plane shear flow is always hydrodynamically
stable under a {\it pure} vertical perturbation.
(2) When $\Omega_3=1/2$, $\sigma=\sqrt{\zeta\epsilon}$. Therefore, the constant angular
momentum accretion flow is always hydrodynamically unstable.
The energy growth rate of perturbation increases with the strain rate i.e.
the eccentricity of the flow.
(3) When $\Omega_3=2/3$, $\sigma=\sqrt{\zeta\epsilon-(4-3\zeta)/9}$.
Therefore, a Keplerian flow with elliptical streamlines
gives rise to unbounded growth, at least in certain time interval when growth
and growth rate due to primary perturbation is very small,
under a {\it pure} vertical perturbation, only if $\zeta>1/3$.

\begin{figure}
\epsscale{1.0}
\plotone{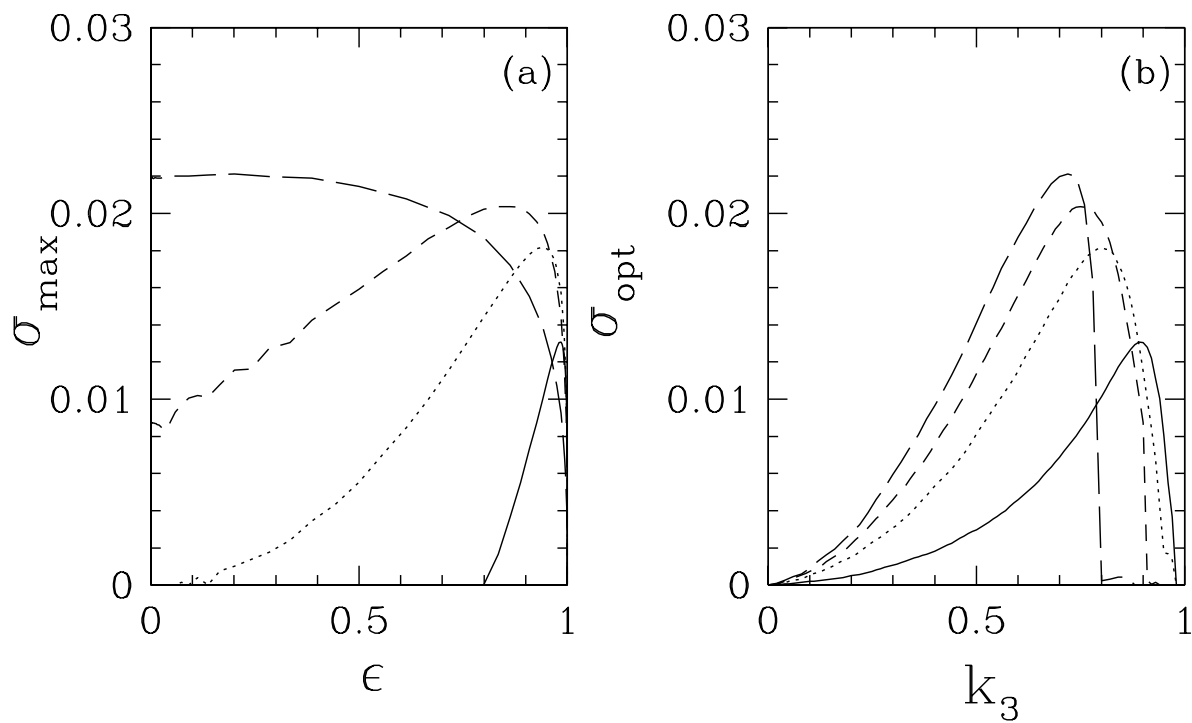}
\vskip2.5cm
\caption{ \label{kepcon}
(a) Variation of maximum growth rate as a
function of eccentricity parameter when solid, dotted, dashed and
long-dashed curves indicate the results for $\zeta=0.2,0.1,
0.05,0.01$ respectively.
(b) Variation of optimum growth rate
as a function of vertical component of perturbation wave-vector
when various curves are same as of (a). $q=1.5$, $|\vec{k}_0|=1$ throughout.
}
\end{figure}

However, there are some other three-dimensional perturbations \footnote{By vertical perturbation
we mean that only the vertical component of initial perturbation wave-vector is
non-zero,
while any other perturbation with a non-zero vertical component of initial wave-vector
is called three-dimensional perturbation.}
which can generate positive growth rate, $\sigma$,
in a Keplerian flow with $\zeta < 1/3$, that we describe by
numerical solutions.
As primary perturbation evolves with time,
eccentricity decreases, and then the energy growth rate due to
secondary perturbation changes.
Figure \ref{kepcon}a shows the
variation of maximum growth rate, $\sigma_{max}$, as a function of
eccentricity parameter, $\epsilon$ \footnote{Note that $\epsilon$ is a parameter 
that carries the information of eccentricity of the system but is not the 
eccentricity itself.}. By ``maximum'' we refer the
quantity obtained by maximizing over the vertical component of
wave-vector, $k_{3}$.
Clearly, for $\zeta < 1/3$, growth rate maximizes for
three-dimensional perturbations with $k_3<1$. At small $\epsilon$
and large $\zeta$, the streamlines of the flow essentially become circular
(see eqn. (\ref{axy})), and thus the growth rate severely decreases 
due to lack of significant elliptical vortex. On the other hand, when $\epsilon$
and $\zeta$ both are small, the background structure reduces to plane shear
and therefore any growth arises due to primary perturbation only.

Figure \ref{kepcon}b shows the variation of optimum growth rate,
$\sigma_{opt}$, as a function of $k_3$.
By ``optimum'' we refer the quantity obtained by maximizing over
$\epsilon$. Interesting fact to note is that the optimum growth rate is
always obtained for three-dimensional perturbation with
significant vertical component.
Moreover, as $\zeta$ increases, the best growth rate is obtained 
at high $\epsilon$ with large $k_3$.
Therefore, three-dimensional growth is more prompt
at larger $\zeta$.

\section{Computation of growth with simultaneous evolution of both perturbations}
Above results verify that at some parameter range the
three-dimensional growth rate due to
secondary perturbation in rotating shear flow is expected to be real and positive, 
which motivates us to analyze the simultaneous evolution of
both perturbations. As primary perturbation evolves, $k_x$
varies with time and therefore $\epsilon$ and then $\bf A$ does so. Thus, 
although eqn. (\ref{keq}) remains still valid, in general
the wave-vector of secondary disturbance is not periodic and the solution of
eqn. (\ref{vmat}) can not be expressed exactly by Floquet modes. 
Therefore, to compute 
growth in energy, one has to find out the elements of energy evolution matrix
\begin{equation}
{\cal M}_{ik}(t)=M_{im}(t)^T M_{mk}(t)
\label{m2}
\end{equation}
whose largest eigenvalue is growth in energy at time $t$. As $M_{mk}(t)$ can be obtained
from eqn. (\ref{matevo}), computation of ${\cal M}_{ik}(t)$ is a trivial job. 
The actual time variance of $\epsilon(t)$ in eqn. (\ref{matevo}) 
(and in eqn. (\ref{mat})) is now considered. Clearly,
${\cal M}_{ik}(t)$ is the instantaneous energy of perturbation of the flow, while we can remind that
$M_{im}(t)$ is the instantaneous velocity of perturbation.

\begin{figure}
\vskip-2cm
\hskip-2.0cm
\includegraphics[width=1.2\columnwidth,angle=0]{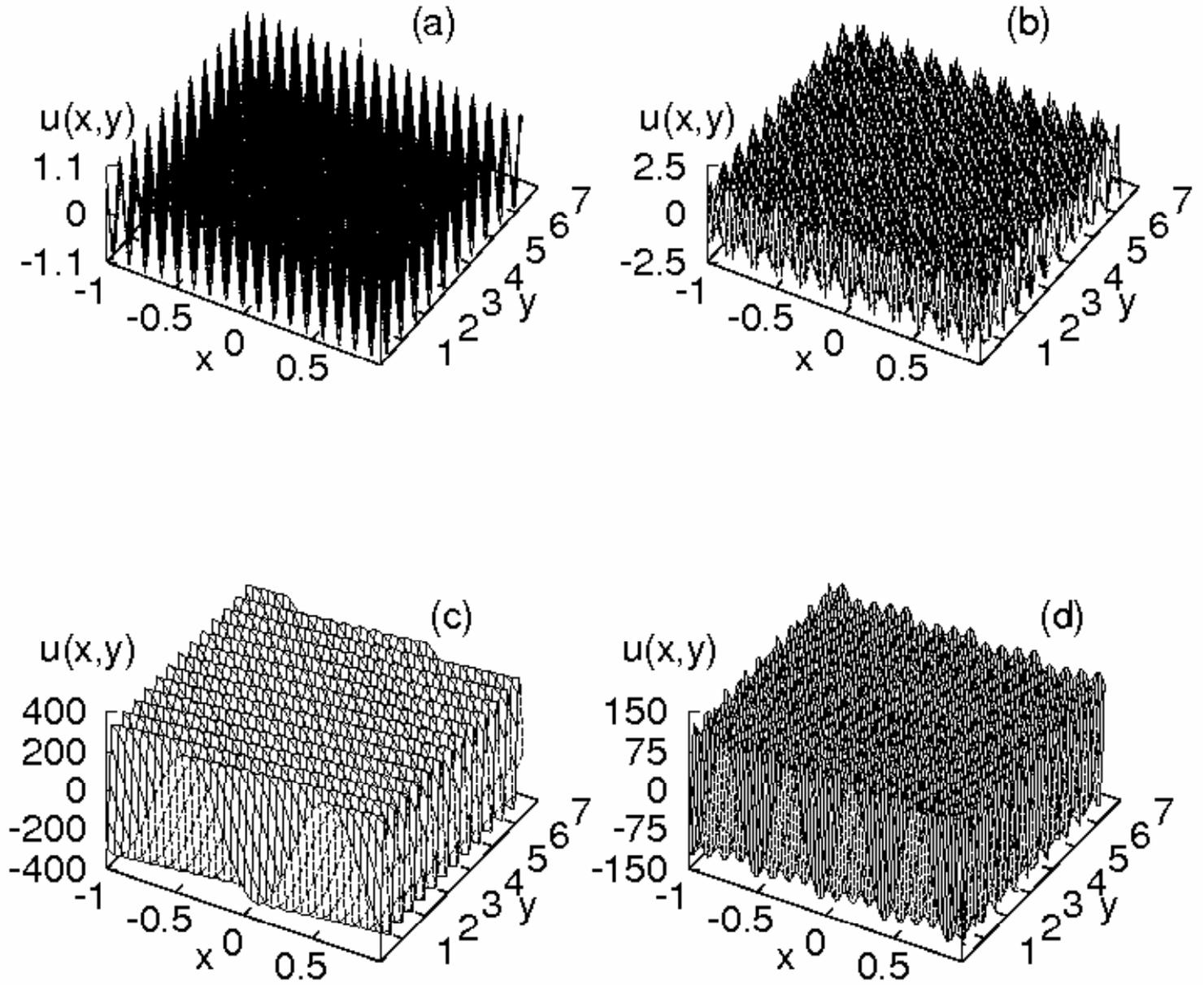}
\caption{ \label{pert}
Evolution of secondary perturbation in the x-y plane for $\zeta=0.1$, when
(a) $t=0$, (b) $t=t_{\rm max}/2=386$, (c) $t=t_{\rm max}=772$, 
(d) $t=3t_{\rm max}/2=1158$. Other parameters
are $k_{x0}=k_{10}/10=-1000$, $k_y=k_{20}/10=1$, $k_3=10$,
and $q=1.5$. The perturbation is normalized by the magnitude of that at $t=0$.
}
\end{figure}

Figure \ref{pert} depicts the evolution of best growing secondary perturbation.
It is clear that perturbation at $t=0$ is a leading wave with
a large negative $k_1$ (as well as $k_x$, although $k_x<<k_1$) and therefore the flow is
highly eccentric at the beginning. With time, $k_1$ (as well as $k_x$) decreases in
magnitude and finally becomes zero at $t=t_{\rm max}$ when growth maximizes.
With further increase of time, the wave becomes trailing and growth starts to decrease.

\begin{figure}
\epsscale{0.9}
\plotone{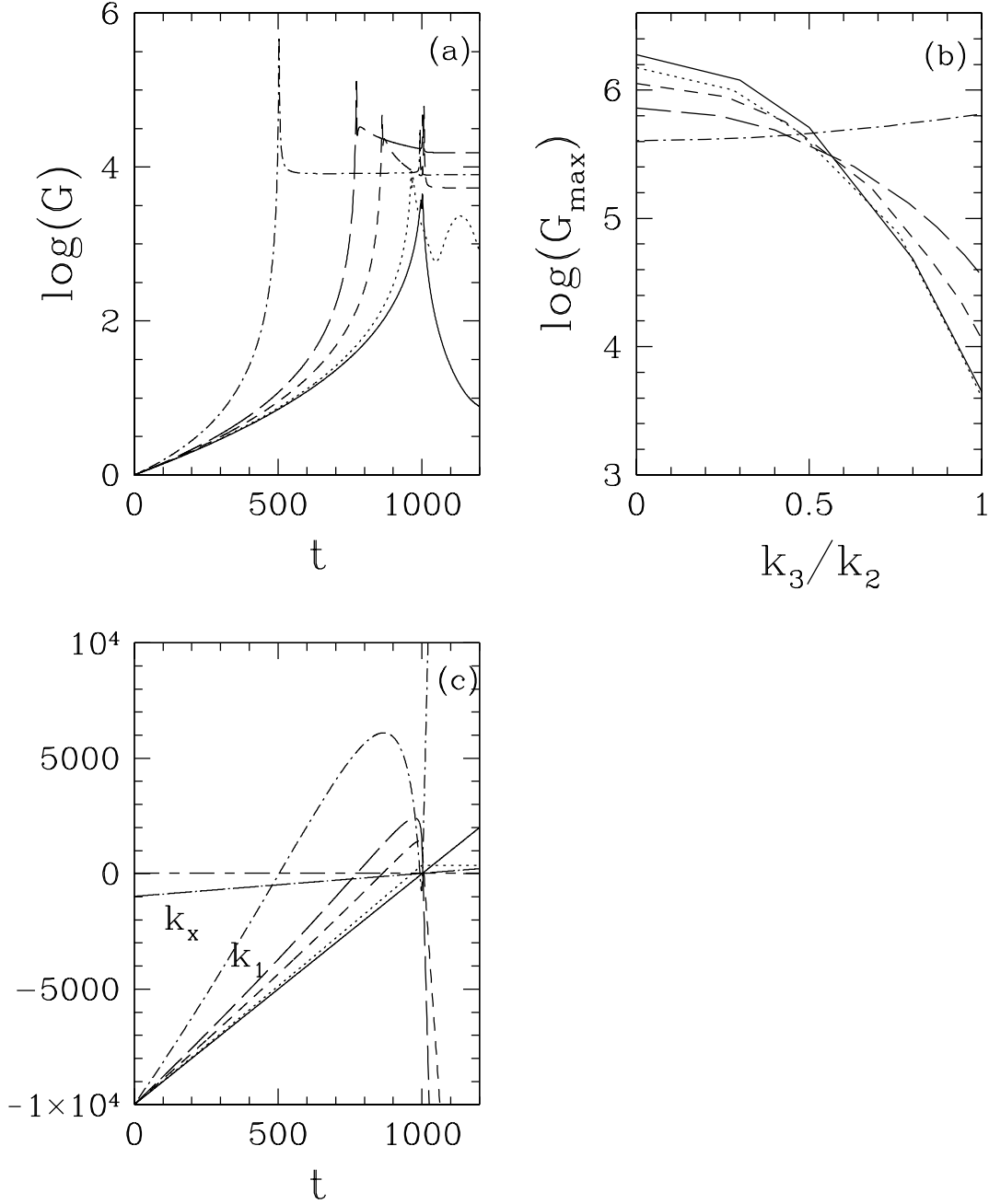}
\vskip2.5cm
\caption{ \label{grow}
(a) Variation of growth as a
function of time due to simultaneous evolution of both perturbations,
when $k_3=10$. Solid, dotted, dashed, 
long-dashed and dot-dashed curves indicate the results for $\zeta=0,0.01,
0.05,0.1,0.4$ respectively.
(b) Variation of maximum growth
as a function of $k_3/k_2$,
when various curves are same as of (a). (c) Variation of radial component of 
primary ($k_x$; dot-long-dashed curve) and secondary ($k_1$; curves have same meaning as
of (a)) perturbation as functions of time, of the cases in (a). The dashed-long-dashed curve 
indicates the line with zero value of the wave vector.
Other parameters are
$k_{x0}=k_{10}/10=-1000$, $k_y=k_{20}/10=1$, and $q=1.5$. Note the change in behavior of curves,
particularly in (b), when $\zeta>1/3$.
}
\end{figure}

Figure \ref{grow}a shows that as $\zeta$ increases, the (first) peak value of growth increases and
that occurs at an earlier time. Comparing with the variation of $k_1$ as a function of $t$,
as shown in Fig. \ref{grow}c, it is very clear that a peak in growth 
appears when $k_1$ approaches
to zero. As $k_1$ in the cases with $\zeta=0.05$ and $0.1$ becomes zero twice, 
corresponding growth maximizes twice too. The second peak appears at $t\sim 1000$ when 
$k_x\sim k_1\rightarrow 0$. Moreover, for $\zeta=0.4$, $k_1$ becomes zero thrice 
(it cuts the zero line at $t\sim 1000$ and becomes negative, but 
immediately turns up and cuts the zero
again). Therefore, the corresponding growth curve attains two peaks 
at $t\sim 1000$ very close to each other, apart from the first one at $t=503$. 
The maximization of growth at the minimization of the radial component of perturbation
wave vector was explained in MAN. This is essentially due to the fact that 
$G\propto 1/k^2$.
If $\zeta=0$, then $k_1$ and $k_x$ both become zero simultaneously as shown in Fig. \ref{grow}c. 
However for $\zeta>0$, $k_1$ increases faster, as follows from
eqn. (\ref{keq}), and becomes zero earlier than $k_x$. 
The interesting fact to note is that underlying growth,
although apparently is of transient kind as shown in Fig. \ref{grow}a, diverges
asymptotically 
for any $\zeta >1/3$, while converges for $\zeta \le 1/3$.
The asymptotic divergence of growth is similar to the instability
one obtains in linear perturbation analysis 
for Poiseuille flow at $R\gsim 5772$ (e.g. Reddy \& Henningson 1993).
The significant asymmetry around $t=t_{\rm max}$ and non-zero asymptotic value in growth curves 
are due to the vertical structure.
Figure \ref{grow}b shows the variation of peak
growth as a function $k_3/k_2$. The quantity $k_3/k_2$ carries
information of how three-dimensional the flow is. 
It is interesting to note that the maximum growth for $\zeta >1/3$ is unbounded so that
increases with vertical structure, while it is bounded for $\zeta \le 1/3$. We know that
in a two-dimensional flow, the maximum growth, $G_{max}$, scales
with $k_{x0}^2$ (MAN). However for $\zeta>0$, $G_{max}$
decreases at small $k_3/k_2$ but increases at large $k_3/k_2$, compared to that for $\zeta=0$.
At around $k_3/k_2=0.5$, which corresponds to the marginally 
two-dimensional perturbation, growth due to secondary perturbation is
comparable to that of primary perturbation. This verifies that secondary perturbation
can govern significant transient growth in a geometrically thin accretion disk with finite 
vertical thickness. However, in three-dimension, when
$k_3/k_2\sim 1$, the secondary perturbation effect always dominates over the primary one.
This indicates that three-dimensional secondary perturbation
enhances energy growth and thereafter any possible non-linear feedback effects
with its elliptical base state.
As $k_{x0}$ and $k_{10}$ ($\sim R^{1/3}$, see UR, MAN)
increase, the vertical structure plays more effective roles to govern growth. When $k_{x0}=k_{10}/10=-10^3$
and $k_y=k_{20}/10=1$, $G_{max}\sim 4\times 10^4$ at $k_3/k_2=1$ for $\zeta=0.1$, which
is an order of magnitude larger compared to that for $\zeta=0$.
If we consider a smaller $R$ with $k_{x0}=k_{10}/10=-10^2$, then
$G_{max}$ at $k_3/k_2=1$ decreases to $\sim 2\times 10^3$ for
$\zeta=0.1$, which is still larger by a factor of two compared to
that for $\zeta=0$. 
Therefore, three-dimensional effects efficiently enhance growth and then presumably 
help to trigger turbulence in shear flows.

\section{Discussions}

We find that significant energy growth of perturbation is possible to govern in shear flow with
the Coriolis force. This system is an idealized local analog of an accretion disk which,
under secondary perturbation, 
definitely exhibits
three-dimensional large growth of transient kind and, in addition, 
sometimes exhibits unbounded late-time growth
at a large amplitude of primary perturbation.
We have explicitly demonstrated
the perturbation effects one by one. First, primary two-dimensional perturbation induces 
vortex into the flow that can be locally seen as elliptical streamlines. This system, which
does not have any exponential growing eigenmode but does exhibit significant transient growth,
has been extensively studied already (e.g. UR; Yecko 2004; MAN).
In this situation, a plane wave perturbation,
that is frozen into the fluid, is sheared along with the
background flow. At $t=0$, the effective wave vector of
perturbation is in the $x$ direction ($k_x>>k_y$) and
is negative which provides very asymmetric leading waves. Therefore, the flow
at this stage is highly eccentric.  As time goes on, the
wavefronts are straightened out by the shear and $|k_x|$
decreases and transient growth increases. When $k_x\sim 0$, i.e. the wavefronts become almost radial,
transient growth is maximum. At this time, the streamlines of
the flow are almost circular. At yet later
time, growth decreases and the wave becomes trailing.
Then it has been argued that if the maximum growth
exceeds the threshold for inducing turbulence,
then this mechanism could drive the system to a turbulent state.
Presumably, once the system becomes turbulent, it remains turbulent
as a result of nonlinear interactions and feedback among the
perturbations. It can be reminded that our present aim is to understand and establish the origin of
viscosity in the flow that must be due to turbulence. The
transfer of mass inward and angular
momentum outward in an accretion disk
is difficult to explain in absence of turbulence. However, the accretion disk
is quite a complex system with, possibly, continuous perturbed flow. Therefore, if we can address
a mechanism to govern turbulence, then its recycling is not a difficult job.

Second, we consider further perturbation, namely secondary perturbation, into the
flow described above. The primarily perturbed shear flow serves as a background for secondary
perturbation whose eccentricity naturally varies with time due to the evolution
of primary perturbation. In this paper, we have especially
demonstrated the evolution of this secondary perturbation which exhibits three-dimensional large 
transient growth in a local Keplerian accretion disk. While primary perturbation itself can produce
large transient growth at a high Reynolds number that might drive the non-linear effects into 
the system, the best perturbation responsible for this effect is two-dimensional. However, it is understood that
perturbations must ultimately decline to zero in presence of viscosity (see e.g. UR),
unless three-dimensional effects are invoked. Therefore, we have addressed
the possible origin of three-dimensional effects that shows a clear route to three-dimensional
hydrodynamic growth and then possible non-linear feedback and turbulence in accretion flows. 
Underlying
growth arises due to the elliptical vortices present in the background,
rather than due to the plane shear which exhibited growth under primary perturbation.

In the standard fluid literature, the elliptical instability has been widely
discussed as a possible route to self-sustained turbulence in linearly perturbed shear flows,
as mentioned in \S1. However, usual emphasis of those investigations is on non-rotating
flows. Craik (1989), while
discussed about the elliptical instability in rotating shear flows, did not focus
on a Keplerian flow, what is of astrophysical interest. Therefore, in the present paper,
we have first discussed the growth rate in standard elliptical flows and
compared the results for
non-rotating flows with that of rotating ones in \S3.1. We have shown that the
growth rate in a Keplerian flow with
constant elliptical streamlines is significantly large compared to that in a non-rotating flow,
particularly at high eccentricity.

However, in reality, when a small section of a Keplerian accretion
disk is considered under a two-dimensional linear
perturbation, the flow governs distorted elliptical streamlines whose structure vary with time.
Therefore, the growth rate due to secondary perturbation at a fixed $\epsilon$ (instantaneous
growth rate) as described in \S3.2 decreases much compared to
that in the flow with idealized elliptical streamlines,
unless the amplitude of primary perturbation, $\zeta$, is very large.
At a $\zeta>1/3$, a pure vertical secondary perturbation produces the best growing eigenmode.
On the other hand, at a $\zeta\le 1/3$,
the best growing eigenmode arises due to other three-dimensional perturbations with a
significant, but not sole, vertical effect. Although the instantaneous growth rate appears
to be small for a small $\zeta$ 
(which is of particular interest), at least compared to 
the case with idealized elliptical streamlines,
actual growth which is the result of simultaneous
evolution of both perturbations as described in \S4 can be large enough to 
exhibit non-linear effects
if the time-scale for the evolution of perturbation is large. The time for the evolution
of perturbation scales
with the Reynolds number of the flow as $R^{1/3}$ (MAN). As perturbation evolves, 
$k_x$ varies from $-\infty$ to $0$ and thus the eccentricity of the flow decreases from
$1$ to $0$. Most of the important three-dimensional
growing modes are generated at the high eccentricity regime when $0.995\le\epsilon
\le 1$ and therefore $10\le |k_x|\le \infty$. Important fact to note is that growth maximizes
for three-dimensional perturbation with a significant vertical effect.

We therefore conclude with an important caveat. UR already showed via
two-dimensional simulations that chaotic motions can persist for a time much longer than
the time scale needed for linear growth. However, the corresponding vorticity decays unless
the vertical structure is there. In the present paper, 
we have shown the existence of three-dimensional
perturbation effects and corresponding eigenmodes which governs large
energy growth and then suggests the possible existence of non-linear effects and self-sustained 
hydrodynamic turbulence in accretion disks. Now one will have to verify our suggestion 
by numerical simulation with proper resolution and possibly also by nonlinear
analytic asymptotic methods.

\begin{acknowledgements}

The author is grateful to Ramesh Narayan for suggesting this problem and for
extensive discussion and encouragement
throughout the course of the work. The author is also thankful to the referee for
his/her constructive suggestions that help to improve the presentation of the paper.
This work was supported in part by NASA grant NNG04GL38G and NSF grant
AST 0307433.

\end{acknowledgements}

{}
\clearpage

\end{document}